\documentclass[useAMS,usenatbib]{mn2e}
\usepackage{times} 
\usepackage{amsmath}
\usepackage{rotating}
\usepackage{amssymb}

\def\aap{A\&A}
\def\apj{ApJ}

\def\apjl{ApJ}
\def\mnras{MNRAS}

\def\apjs{ApJS}
\def\prd{Phys. Rev. D}

\newcommand{\tvec}[1]{\underline{#1}}
\newcommand{\svec}[1]{\overline{#1}}

\newcommand{\ssvec}[1]{\overline{\overline{#1}}}

\newcommand{\beq}{\begin{equation}}
\newcommand{\eeq}{\end{equation}}
\newcommand{\beqn}{\begin{eqnarray}}
\newcommand{\eeqn}{\end{eqnarray}}
\newcommand{\lb}{\left}
\newcommand{\rb}{\right}
\newcommand{\diff}[2]{\frac{\mathrm{d} #1}{\mathrm{d} #2}}

\newcommand{\partdiff}[2]{\frac{\partial #1}{\partial #2}}
\newcommand{\partddiff}[2]{\frac{\partial^2 #1}{\partial #2^2}}

\newcommand{\dint}[1]{\mathrm{d}#1}

\newcommand{\nn}{\nonumber}

\title[Dark Matter Caustics]
{Dark Matter Caustics}
\author[White \& Vogelsberger] 
{
Simon D.M. White\thanks{swhite@mpa-garching.mpg.de},
Mark Vogelsberger\thanks{vogelsma@mpa-garching.mpg.de} \\
Max-Planck Institut fuer Astrophysik,
Karl-Schwarzschild Strasse 1, 
D-85748 Garching, 
Germany 
}

\begin{document}
\date{Accepted ???. Received ???; in original form ???}

\pagerange{\pageref{firstpage}--\pageref{lastpage}} \pubyear{2008}

\maketitle

\label{firstpage}

\begin{abstract}
Caustics are a generic feature of the nonlinear growth of structure in
the dark matter distribution. If the dark matter were absolutely cold,
its mass density would diverge at caustics, and the integrated
annihilation probability would also diverge for individual particles
participating in them.  For realistic dark matter candidates, this
behaviour is regularised by small but non-zero initial thermal
velocities. We present a mathematical treatment of evolution from Hot,
Warm or Cold Dark Matter initial conditions which can be directly
implemented in cosmological N-body codes. It allows the identification
of caustics and the estimation of their annihilation radiation in fully
general simulations of structure formation.
\end{abstract}


\section{Introduction}

The idea that the dark matter might consist of a collisionless ``gas'' of
weakly interacting, neutral particles was first published by
\cite{1973ApJ...180....7C} and \cite{1976A&A....49..437S}, though earlier
discussion can also be found in the textbook of \cite{Zeldovich1971}. 
These authors proposed neutrinos as a promising dark matter candidate,
and a claimed measurement of the electron neutrino mass at very nearly the
expected value \citep{1981Lyubimov} led to a flurry of interest in
neutrino-dominated, so-called ``Hot Dark Matter'' universes.  This continued
until detailed numerical simulations showed the predictions for low-redshift
structure to be quite inconsistent with observation
\citep{1983ApJ...274L...1W}. Attention then shifted towards more exotic
particle candidates for Warm or Cold Dark Matter \citep{1982PhRvL..48..223P,
1982ApJ...263L...1P}.

If a cold collisionless gas evolves from near-uniform initial
conditions under the influence of gravity, the nonlinear phases of
growth generically involve caustics analogous to those formed when
light propagates through a non-uniform medium. This connection was
explored in some depth by Russian cosmologists interested in
neutrino-dominated universes 
\citep{1982GApFD..20..111A,1983SvPhU.139..153Z}. 
Caustics are also a very evident
feature of the similarity solutions for cold spherical infall
published by \cite{1984ApJ...281....1F} and \cite{1985ApJS...58...39B}. It
was another 15 years, however, before \cite{2001PhRvD..64f3515H} realised that dark
matter annihilation could be very substantially enhanced in such
caustics.  He showed that for absolutely cold dark matter the
annihilation probability diverges logarithmically as a particle passes
through a caustic, and that for realistic dark matter candidates this
divergence is tamed by the small but finite initial thermal velocities
of the particles. He argued that the annihilation radiation from dark
halos might be dominated by emission from caustics.

Twenty years earlier \cite{1982PAZh....8..259Z} had noted that thermal
velocities limit the densities achievable in dark matter caustics, and
the first rigorous calculation of annihilation rates in caustics was
carried out by \cite{2006MNRAS.366.1217M} for Bertschinger's (1985)
similarity solution.  They found caustics to enhance the total
annihilation flux substantially in the outer regions for plausible
values of the initial dark matter velocity dispersion, but to be
progressively less important at smaller radii.  It is unclear whether
either of these results will apply in general, since the behaviour of
the similarity solution is strongly influenced by its spherical
symmetry (which reduces its phase-space dimensionality from 6 to 2)
and by its lack of small-scale structure.

In the present paper we present a theoretical treatment of the growth
of structure which shows how the geodesic deviation equation can be
used to follow local phase-space structure in a Lagrangian treatment
of nonlinear evolution. This formalism is well suited for
implementation in N-body simulation codes, allowing the annihilation
signal from caustics to be treated in full generality provided
numerical artifacts from discretisation and integration error can be
kept under control. We have presented results from a first
implementation of this scheme in \cite{2008MNRAS.385..236V}. 
A closely related but somewhat more complex scheme is 
described by \cite{2005MNRAS.359..123A}. Here we
complement this work by giving a fuller description of the mathematics
behind the approach, in particular of the regularisation of caustic
densities by the finite velocity dispersion of the dark matter. In the
next section we describe our idealisation of the initial conditions
for structure formation in WIMP-dominated cosmologies. Section 3 then
presents useful general results for nonlinear evolution from these
initial conditions. These are used in section 4 to describe the
evolution of the local phase-space structure, in particular of its
projection onto configuration space, following individual particle
trajectories.  Caustic passages can be identified and the associated
annihilation signal can be calculated explicitly. A final section
discusses possible future uses of this approach.

\section{Idealised initial conditions for structure formation}

In the current standard paradigm for cosmological structure formation, the
dark matter is assumed to be a weakly interacting massive particle which
decoupled from all other matter and radiation fields at an early epoch, well
before the universe became matter-dominated. Since this time, the dark matter
has interacted with other components only through gravity. In most such models
there is a period after the transition to matter-domination when density
fluctuations in all components are linear on all scales, and the residual
thermal velocities of the dark matter particles are small compared to the
large-scale velocities induced by density inhomogeneities. In this paper we
will take an idealised representation of this situation as the initial
condition for later evolution of the dark matter distribution.

Thus we assume that at the initial time $t_0$ the phase-space
density of dark matter particles can be written as
\beq
f(\tvec{q},\tvec{p},t_0) = \rho(t_0)/m_p~ (1+\delta(\tvec{q})) ~ \mathcal{N}(
(\tvec{p} - \tvec{V}(\tvec{q}))/\sigma) , 
\eeq
where $\rho(t)$ is the (time-varying) mean mass density of dark
matter, $m_p$ is the dark matter particle mass, $\tvec{q}$ and
$\tvec{p}$ are position and velocity at the initial time and will be
used as Lagrangian coordinates labelling individual dark matter
particles as they follow their trajectories, $\delta(\tvec{q})$ is the
initial linear overdensity at position $\tvec{q}$,
$\tvec{V}(\tvec{q})$ is the linear peculiar velocity at $\tvec{q}$
and, for growing mode linear perturbations in the matter-dominated
regime, is related to $\delta(\tvec{q})$ through
\beq
\tvec{V}(\tvec{q}) = -\nabla\phi\, ~~~~~\delta(\tvec{q})\propto\nabla^2\phi , 
\eeq 
where $\phi(\tvec{q})$ is the gravitational potential generated by
the linear fluctuation field, $\sigma$ is the thermal velocity dispersion
of the dark matter particles at the initial time, and $\mathcal{N}$ denotes the standard
normal distribution in three dimensions. Our assumption that the initial
density field is linear implies that $\langle \delta^2\rangle \ll 1$, while
our assumption that the initial dark matter distribution is cold implies
that $\sigma^2 \ll \langle |\tvec{V}|^2\rangle$. Note that the latter condition
applies even in the hot dark matter cosmology, provided the initial time
$t_0$ is taken sufficiently late.
 
In phase-space the dark matter is thus confined initially very close to the
three-dimensional ``sheet'' $\tvec{p} = \tvec{V}(\tvec{q})$ and, indeed, its
distribution becomes exactly three-dimensional in the cold limit
$\sigma \rightarrow 0$. The projection of this sheet onto configuration space
(the three-density) is very nearly uniform.  It proves useful to work in
initial coordinates where the velocities at each point are taken relative to
$\tvec{V}(\tvec{q})$. We therefore define
\beq
\tvec{A}(\tvec{q},\tvec{p}) = \tvec{p} - \tvec{V}(\tvec{q}), 
\eeq
and we approximate the initial phase-space density of the dark matter as
\beq
f(\tvec{q},\tvec{A},t_0) = f_0 ~ \mathcal{N}(\tvec{A}/\sigma),
\label{eq:InitialCondition}
\eeq
where we neglect the spatial modulation by the factor $(1+\delta)$ so that
$f_0 =\rho(t_0)/m_p$ and the phase-space density becomes independent of
$\tvec{q}$. Because $\tvec{V}$ is the gradient of a scalar field the
second-rank tensor $\partial \tvec{A} / \partial \tvec{q}$ is symmetric and the
transformation of variables $(\tvec{q},\tvec{p})\leftrightarrow
(\tvec{q},\tvec{A})$ is canonical \citep{2008gady.book.....B}.  This will
important below.

\section{Evolution of the dark matter distribution}

We will assume that dark matter particles interact purely gravitationally at
all times of interest. Each then moves independently of the others subject
only to the collective gravitational potential. Particle accelerations depend
on position and time, but not on velocity, and trajectories can be derived
from the time-dependent Hamiltonian of the system \citep[e.g.][]{1980lssu.book.....P}. The
evolution of the dark matter distribution is thus a Hamiltonian flow. A
thorough development of the properties of such flows can be found in Appendix
D of \cite{2008gady.book.....B}.  We characterise the phase-space position of
a particle at time $t$ by its position $\tvec{x}$ and its peculiar
velocity $\tvec{v}$. In a Hamiltonian flow, particle trajectories through
phase-space never intersect, so we can write the phase-space position at time
$t$ as a unique and invertible function of the initial phase-space position,
i.e.
\beq
\tvec{x} \equiv \tvec{x}(\tvec{q},\tvec{p}, t),~~~ \tvec{v} \equiv \tvec{v}(\tvec{q},\tvec{p}, t) \nn
\eeq
with the well-defined inverse relation
\beq
\tvec{q} \equiv \tvec{q}(\tvec{x},\tvec{v}, t),~~~\tvec{p} \equiv \tvec{p}(\tvec{x},\tvec{v}, t). \nn
\eeq
This is a standard Hamiltonian map so the transformation
$(\tvec{q},\tvec{p}) \leftrightarrow (\tvec{x},\tvec{v})$ which it defines is
canonical, incompressible and symplectic 
\citep[see][for the mathematical significance of these terms]{2008gady.book.....B}.
As noted above, the transformation of variables 
$(\tvec{q},\tvec{p}) \leftrightarrow (\tvec{q},\tvec{A})$ is canonical, 
so we can restate the (canonical) relation between initial and final 
configurations as
\beq
\tvec{x} \equiv \tvec{x}(\tvec{q},\tvec{A}, t),~~~\tvec{v} \equiv \tvec{v}(\tvec{q},\tvec{A}, t) \nn
\eeq
with 
\beq
\tvec{q} \equiv \tvec{q}(\tvec{x},\tvec{v}, t),~~~\tvec{A} \equiv
\tvec{A}(\tvec{x},\tvec{v}, t). \nn
\eeq
This simplifies the description of our problem considerably.
\begin{figure}
\centerline{\includegraphics[width=1.0\linewidth]{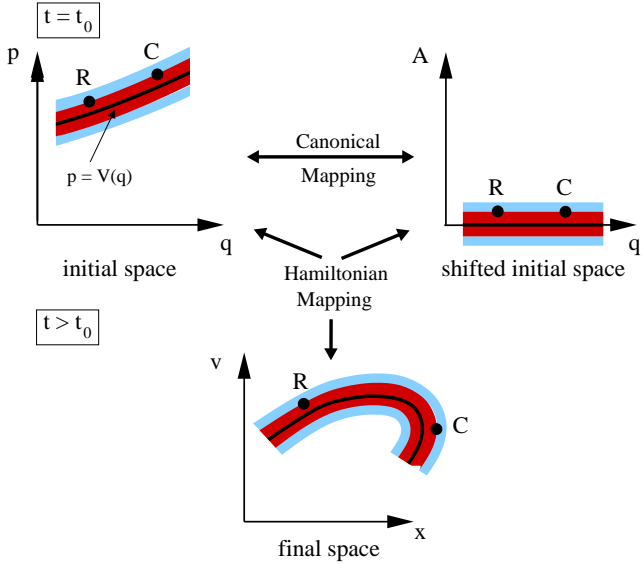}}
\caption{Illustration of our calculations for an analogous
  1-dimensional system. At the initial time $t_0$, the dark matter
  phase space density is non-zero only in regions of phase space close
  to the central sheet $p=V(q)$, indicated by the solid curve in the
  upper left diagram. The shaded regions surrounding this curve
  indicate the 1 and 2-$\sigma$ regions of the (Gaussian) phase space
  density distribution. For convenience, we change the initial
  velocity coordinate in phase space to $A=p-V(q)$. As shown in the
  upper right diagram, the equidensity contours of the phase space
  density then correspond to $A={\rm const.}$, and the maximum phase
  space density occurs on the $A=0$ axis. Dynamical evolution distorts
  these initial phase-space distributions according to the Hamiltonian
  flow $(q,p) \leftrightarrow (x,v)$, producing a phase space
  distribution at a later time $t$ which is indicated schematically in
  the lower diagram. The collisionless Boltzmann equation guarantees
  that the phase space density is preserved by this map, as indicated
  by the shaded regions. Our plots indicate how two points R and C,
  corresponding to two different dark matter particles, are
  transformed by these maps. The (1-dimensional) space density in the
  neighborhood of each is given by projecting the phase space density
  down the velocity axis, and is uniform in the initial space. At time
  $t$, R is at a regular point of its trajectory ($\partial A/\partial v
  \neq 0$), whereas C is passing a caustic ($\partial A/\partial v
  =0$). Clearly the space density at R depends on the local slope of
  its $A= {\rm const.}$ line, while at $C$ it depends on the curvature
  of this line and on the offset of C from the central sheet of its
  stream.}
\label{Fig:SchematicMapping}
\end{figure}

The collisionless Boltzmann equation is an immediate consequence of
the incompressibility (in 6-D) of Hamiltonian flows -- phase-space
density is conserved along every trajectory in the flow. 
Eq.~(\ref{eq:InitialCondition}) then provides a complete formal solution
for the evolution of the phase-space density distribution of the dark
matter distribution:
\beq
f(\tvec{x}, \tvec{v}, t) =  f_0 ~ \mathcal{N}(\tvec{A}(\tvec{x}, \tvec{v}, t)/\sigma). 
\label{eq:GeneralSolution}
\eeq
The maximum phase-space density at time $t$ is $f_0$, and this density
is achieved everywhere on a 3-dimensional subspace defined implicitly
by $\tvec{A}(\tvec{x}, \tvec{v}, t) = 0$. Phase-space densities are
only significantly different from zero at points which are
sufficiently close to this subspace which we refer to below as the
``central sheet'' of the phase-space distribution. The geometry of
this sheet is very simple at early times: its projection onto 3-space
is (approximately) uniform and only one point of the sheet projects
onto each $\tvec{x}$-position. Non-linear evolution stretches and
folds the sheet, but does not tear it. It can then pass through a
given $\tvec{x}$-position multiple times, producing a series of
streams, each with a different velocity $\tvec{v}$.  Caustics arise on
the boundaries between regions with different numbers of
streams. Figure 1 illustrates this situation for an analogous
1-dimensional system.

It is important to realise that the solution of 
Eq.~(\ref{eq:GeneralSolution}) depends {\it only} on the assumed initial condition
and on the fact that the dark matter obeys the collisionless Boltzmann
equation. Thus it holds in the absence of any symmetry and during strongly
non-equilibrium phases of evolution.  In addition, it does not assume the
gravitational potential to be generated by the dark matter alone, so it is
valid, for example, in the inner regions of galaxies, where the gravitational
effects of the baryonic components are dominant. These influence the
trajectories of individual dark matter particles, and so the details of
$\tvec{A}(\tvec{x}, \tvec{v}, t)$, but they do not affect the Hamiltonian
nature of the flow or the validity of Eq.~(\ref{eq:GeneralSolution}).

The Hamiltonian nature of the flow has a number of consequences for the map
$(\tvec{q},\tvec{A}) \leftrightarrow (\tvec{x},\tvec{v})$. If we define
6-vectors $\svec{w} \equiv (\tvec{q},\tvec{A})$ and $\svec{W} \equiv
(\tvec{x},\tvec{v})$, then the 6-tensors 
\beq 
\ssvec{D}\equiv \partdiff{\svec{W}}{\svec{w}} ~~~ {\rm and} 
~~~ \ssvec{D^\prime}\equiv \partdiff{\svec{w}}{\svec{W}} \nn 
\eeq
 satisfy the relations 
\beq 
\ssvec{D} ~ \ssvec{D^\prime}=\ssvec{D^\prime} ~ \ssvec{D}=\ssvec{I},~~~ {\rm
  det}(\ssvec{D})={\rm det}(\ssvec{D^\prime})=1.  
\eeq 
The first relation merely states that the backward transformation reverses the
forward transformation, so the matrix corresponding to the former is the
inverse of that corresponding to the latter. The second relation states that
both matrices have unit determinant so that the transformations are
volume- and orientation-preserving. The conservation of phase-space density
expressed by the collisionless Boltzmann equation is a consequence of this
second property. Further important properties follow from the fact that these
matrices are symplectic. In particular, the velocity and space parts of the
forward and backward transformations are related by
\beq
\partdiff{\tvec{x}}{\tvec{q}} = \partdiff{\tvec{A}}{\tvec{v}}, \,
\partdiff{\tvec{x}}{\tvec{A}} = -\partdiff{\tvec{q}}{\tvec{v}}, \,
\partdiff{\tvec{v}}{\tvec{q}} = -\partdiff{\tvec{A}}{\tvec{x}}, \,
\partdiff{\tvec{v}}{\tvec{A}} = \partdiff{\tvec{q}}{\tvec{x}}, 
\label{eq:SymplecticRelations}
\eeq
where in each equation the partial derivative on the l.h.s. refers
to the forward map so that the independent variables are
$\tvec{q}$ and $\tvec{A}$, while the partial derivative on the r.h.s.
refers to the reverse map so that the independent variables are
$\tvec{x}$ and $\tvec{v}$. We will use some of these relations later.

\section{Variation of the 3-density along particle trajectories}

The differential annihilation probability for an individual dark
matter particle depends on the (velocity-dependent) annihilation
cross-section $\sigma_\times(v)$ and the local phase-space distribution 
as
\beq
\diff{P}{t}(\tvec{x}, \tvec{v}, t) = \int d^3v^\prime f(\tvec{x}, \tvec{v^\prime},
t) \sigma_\times\lb(|\tvec{v^\prime} - \tvec{v}|\rb)|\tvec{v^\prime} - \tvec{v}|. 
\eeq
In the following we will assume that $\sigma_\times(v)\propto 1/v$ as is the case for
many (but not all) dark matter candidates.  This relation then simplifies to
\beq
\diff{P}{t}(\tvec{x}, \tvec{v}, t) = \langle\sigma_\times v\rangle \int d^3v^\prime
f(\tvec{x}, \tvec{v^\prime}, t) = \frac{\langle\sigma_\times v\rangle}{m_p} ~ \rho(\tvec{x}, t), 
\label{eq:AnnihilationRate}
\eeq
where $\rho(\tvec{x}, t)$ is the local 3-space mass density of dark matter 
and $\langle\sigma_\times v\rangle$ is the thermally averaged 
velocity-weighted annihilation cross-section.  The
total annihilation probability over a finite time interval is thus simply
obtained by integrating the local dark matter 3-density along the particle
trajectory. This density is made up of two distinct components, one due to
particles which are part of the same stream as the particle in question,
and so were its neighbours in the initial conditions, and one due
to particles in other streams, which typically originated in distant
parts of phase-space. Caustics arise in the first of these two components
and so we will concentrate on it in the following.

For our assumed initial condition, each particle can be specified by its
initial phase-space position $(\tvec{q}_p, \tvec{A}_p)$, where $ |\tvec{A}_p|
\sim \sigma$ is very small. The subscript $p$ here identifies that the coordinates
belong to the specific particle under consideration; it has nothing to do
with the initial phase-space coordinate $\tvec{p}$. 
The particle's  later trajectory is then $\tvec{x}_p(t) =
\tvec{x}(\tvec{q}_p,\tvec{A}_p,t)$, $\tvec{v}_p(t) = \tvec{v}
(\tvec{q}_p,\tvec{A}_p,t)$, and the 3-space stream density at its position can
be obtained by integrating the phase-space density over all velocities:
\beq
  \rho_s(\tvec{x}_p(t)) \!= \!\!
  \frac{f_0 m_p}{(2\pi\sigma^2)^{3/2}} \int \dint^3{v} \,
  \exp\lb(-\frac{|\tvec{A}\lb(\tvec{x}_p,\tvec{v},t\rb)|^2}{2\sigma^2}\rb).
\label{eq:StreamDensity}
\eeq
The velocity integral is restricted to velocities such that
$\tvec{q}\lb(\tvec{x}_p,\tvec{v},t\rb)$ remains in the neighborhood of
$\tvec{q}_p$. Since $\sigma$ is very small, we can simplify by carrying out a
Taylor expansion of $ \tvec{A}\lb(\tvec{x}_p,\tvec{v},t\rb)$ around $\tvec{v}
= \tvec{v}_p$,
\beq
\tvec{A}\lb(\tvec{x}_p,\tvec{v},t\rb) = \tvec{A}_p + 
\delta\tvec{v}\partdiff{\tvec{A}}{\tvec{v}}
+ \frac{1}{2} \delta\tvec{v}\partddiff{\tvec{A}}{\tvec{v}} \delta\tvec{v}^T,
\label{eq:TaylorVelocity}
\eeq
where $\delta\tvec{v} = \tvec{v} - \tvec{v}_p$ and the partial derivatives
are evaluated at $(\tvec{x}_p, \tvec{v}_p)$.

\subsection{Densities at regular points}

At almost all points of the particle's trajectory the linear map in
the second term on the r.h.s. of Eq.~(\ref{eq:TaylorVelocity}) is
non-singular (i.e. the determinant of the corresponding matrix is
non-zero). These will be called ``regular points'' of the trajectory
in the following. At such points there exists a small value 
of $\delta\tvec{v}$, say $\delta\tvec{v}_c$, for which the sum of the
first two terms vanishes; $(\tvec{x}_p, \tvec{v}_p +
\delta\tvec{v}_c)$ is then the intersection at time $t$ of $\tvec{x} =
\tvec{x}_p$ with the central sheet of the stream to which our particle
belongs. The integral for $\rho_s(\tvec{x}_p)$ becomes particularly
simple if we centre the velocity integration on this point. To lowest
order, we have
\beq
 |\tvec{A}|^2 = \delta\tvec{v}^\prime\partdiff{\tvec{A}}{\tvec{v}}
\lb(\partdiff{\tvec{A}}{\tvec{v}}\rb)^T  \delta{\tvec{v}^\prime}^T, 
\label{eq:Asquared}
\eeq
where $\delta\tvec{v}^\prime = \tvec{v} - \tvec{v}_p - \delta\tvec{v}_c$.  The
tensor product $(\partial\tvec{A}/\partial\tvec{v})
(\partial\tvec{A}/\partial\tvec{v})^T $ is clearly symmetric and must have
positive eigen-values. Without loss of generality, let us define the principal
axes in velocity space so that these eigen-values are $s_1^2\geq s_2^2\geq
s_3^2 > 0$. The condition that our point be regular forces the final strict
inequality, since an obvious consequence of Eq.~(\ref{eq:Asquared}) is
\beq {\rm det}
\lb(\partdiff{\tvec{A}}{\tvec{v}}\rb) = s_1s_2s_3.  
\eeq 
In this velocity frame the integral in Eq.~(\ref{eq:StreamDensity}) takes 
a simple form which we can easily evaluate,
\beqn
 \label{eq:StreamDensityRegular}  
  \rho_s(\tvec{x}_p(t)) \!\!\!  &=& \!\!\!
  \frac{f_0m_p}{(2\pi\sigma^2)^{3/2}} \int \dint^3{v} \, 
  \exp\lb(-\frac{1}{2\sigma^2}\sum_{i=1}^3 s_i^2\delta {v_i^\prime}^2\rb)   \\ 
  \!\!\!  &=& \!\!\!  \frac{f_0m_p}{|s_1s_2s_3|} 
  =  f_0m_p \lb|{\rm det}\lb(\partdiff{\tvec{A}}{\tvec{v}}\rb)\rb|^{-1}   \nn \\
  \!\!\!  &=& \!\!\! \!\! f_0m_p \lb|{\rm det}\lb(\partdiff{\tvec{x}}{\tvec{q}}\rb)\rb|^{-1}.  \nn
\eeqn
The last equality follows from one of the relations listed in
Eq.~(\ref{eq:SymplecticRelations}) and demonstrates explicitly that the stream
density obtained here is identical to that obtained by forward integration of
the geodesic deviation equation in \cite{1999MNRAS.307..495H} and
\cite{2008MNRAS.385..236V}. This is a manifestation of the fact that the local
velocity distribution in a stream is distorted in a way which exactly mirrors
that of the density field. To lowest order, the map rotates each infinitesimal
Lagrangian volume and then stretches it by different amounts $s_1$, $s_2$ and
$s_3$ along three orthogonal axes. (Note that the $s_i$ can be negative.) The
(initially isotropic) velocity distribution at the central point is compressed
by these same factors along the same set of axes.

\subsection{Densities at caustic crossing}

As a particle moves along its trajectory, it will occasionally pass through
discrete points where the determinant of $\partial \tvec{A} / \partial \tvec{v}$
vanishes and the corresponding linear map becomes singular. At such points at
least one of the stretch factors must vanish, and Eq.~(\ref{eq:StreamDensityRegular}) 
predicts an infinite stream density. These points are the caustics of the map. 
In the following we will neglect higher order singularities, where two or more 
stretch factors vanish, and assume $s_1^2\geq s_2^2>0, s_3^2=0$ at such caustic 
crossings \citep[see][for discussion of higher order singularities]{1999MNRAS.307..877T}.

To simplify the integral in Eq.~(\ref{eq:StreamDensity}) it is useful to
take coordinates in $\tvec{v}$-space along the principal axes of
$(\partial\tvec{A}/\partial\tvec{v}) (\partial\tvec{A}/\partial\tvec{v})^T$. Unit vectors
along the axes corresponding to $s_1^2$ and $s_2^2$ are rotated into a pair of
orthogonal vectors in $\tvec{A}$-space by the linear map
$\partial \tvec{A}/ \partial \tvec{v}$, and in addition are stretched by factors
$s_1$ and $s_2$. We use these rotated directions to define our 1 and 2 axes in
$\tvec{A}$-space. All $\tvec{v}$-vectors are then projected onto the 1-2 plane
in $\tvec{A}$-space. To lowest order, Eq.~(\ref{eq:TaylorVelocity}) becomes
\beqn
\quad \tvec{A}(\tvec{x}_p,\tvec{v},t) 
&=& s_1~\delta v_1^\prime~\tvec{e}_1 + s_2~\delta v_2^\prime~\tvec{e}_2  \nn \\
&+& \lb(A_{p,3} + s_3~\delta v_3 + \frac{1}{2}~\partddiff{A_3}{v_3}~{\delta v_3}^2\rb)\tvec{e}_3,
\label{eq:ACaustic}
\eeqn
where the $\tvec{e}_i$ are unit vectors along the coordinate directions in
$\tvec{A}$-space. We have shifted the origin in $\tvec{v}$-space in order to
simplify the coefficients of $\tvec{e}_1$ and $\tvec{e}_2$. By allowing the
shift to depend on $\delta v_3$, all terms independent of $\delta v_1$ and
$\delta v_2$ can be removed; remaining second-order terms are then small
compared to the linear term which is retained. A similar manipulation is not
possible for the coefficient of $\tvec{e}_3$ because its linear term vanishes
at caustic crossing (i.e. $s_3=0$ when $t=t_c$) and we must retain both
constant and quadratic terms. Note that the {\it only} quadratic term we need
to retain is that involving $\delta v_3$ alone, since all others are
subdominant to the linear terms involving $\delta v^\prime_1$ and $\delta
v^\prime_2$ or are removed by the origin shift made for each value of $\delta
v_3$. Inserting this expression into Eq.~(\ref{eq:StreamDensity}) and setting
$s_3=0$ (i.e. $t=t_c$), the integrals over $v_1$ and $v_2$ can be carried out
as before and we are left with
\beqn
  \rho_s(\tvec{x}_p(t_c)) 
  \!\!\!\!\!&=&\!\!\!\!\!  
  \frac{f_0 m_p}{|s_1s_2|}~~\frac{1}{(2\pi\sigma^2)^{1/2}}  \nn \\
  \!\!\!\!\!&\times&\!\!\!\!\!\! \int\limits_{-\infty}^{+\infty} \dint{v_3} \, 
  \exp\lb(-\frac{1}{2\sigma^2}\lb(A_{p,3} +
  \frac{1}{2}~\partddiff{A_3}{v_3}~{\delta v_3}^2\rb)^2\rb)  \nn  \\ 
  \!\!\!\!\!&=&\!\!\!\!\!\! \frac{f_0 m_p}{|s_1s_2|}~~~\frac{\Gamma^\prime\lb(\pm |A_{p,3}|/\sigma\rb)}{|\sigma
  \partial^2{A_3}/\partial{v_3^2}|^{1/2}}~~\exp (- A_{p,3}^2/2\sigma^2)\,
\label{eq:StreamDensityCaustic}
\eeqn
where the dimensionless function $\Gamma^\prime$ is of order unity and is defined by
\beqn 
\Gamma^\prime(X) = \frac{1}{(2\pi)^{1/2}} \int\limits_{-\infty}^{+\infty}
\dint{y} \, \exp\lb(X^2/2-(X +  y^2/2)^2/2\rb).  
\eeqn 
The sign of the argument of $\Gamma^\prime$ in 
Eq.~(\ref{eq:StreamDensityCaustic}) is positive when $ A_{p,3}$ and
$\partial^2{A_3}/\partial{v_3^2}$ have the same sign and negative
otherwise.  Note that $\Gamma^\prime$ can be expressed in terms of
a modified Bessel function, but we forgo the details here.

As before, the mass density at the particle's position is related to the
initial density $f_0m_p$ by the product of three dilution/compression
factors. The factors associated with the two axes in the plane of the
caustic correspond to those we obtained above for regular points of
the trajectory; caustic formation does not significantly effect the
distortion of the dark matter distribution in directions parallel to
the caustic. The compression/dilution in the direction perpendicular
to the caustic is of different form, depending explicitly on how cold
the initial conditions were (i.e. on the value of $\sigma$) and on how
close the particle is to the central sheet of its stream (i.e. on the
value of $A_{p,3}/\sigma$), as well as on the overall structure of the
Hamiltonian flow, as encapsulated by the second derivative,
$\partial^2{A_3}/\partial{v_3^2}(\tvec{x}_p,\tvec{v}_p,t_c)$. Note
that as the initial conditions are made colder, the maximum density
achieved during caustic passage increases as $\sigma^{-1/2}$.

For particles which lie away from the central sheet of the stream, $A_{p,3}$
is non-zero. If $A_{p,3}$ and $\partial^2{A_3}/\partial{v_3^2}$ have the same
sign, the maximum of $\rho_s(\tvec{x}_p(t))$ then occurs either slightly
before or slightly after $t_c$. This is because a small but non-zero value of
$s_3$ allows the linear term in the coefficient of $\tvec{e}_3$ in
Eq.~(\ref{eq:ACaustic}) to be significant. For one particular $s_3$ the
resulting quadratic has an extremum at $A_3=0$. This value of $s_3$ obtains at
the moment when our particle is spatially coincident with the caustic of the
central sheet of its stream, and so sees the corresponding density (see
Fig.~1). The exact value of the maximum density is not important in practice,
since it appears only in the argument of a logarithm when we estimate the
total annihilation radiation from caustic passages. For simplicity, we will
set $A_{p.3}=0$ when we use Eq.~(\ref{eq:StreamDensityCaustic}) below to
estimate the maximum density during a caustic passage. In this case,
$\Gamma^\prime$ can be expressed in terms of the standard complete
$\Gamma$-function, $\Gamma^\prime(0) = 2^{5/4} \Gamma(5/4) / \sqrt{\pi}$.

\subsection{Integrating annihilation rates through caustics}

The formalism developed above is particularly powerful when embedded
in a high-resolution simulation of cosmic structure
formation. \cite{2008MNRAS.385..236V} showed how the geodesic deviation
equation can be integrated in parallel with the $N$-body equations of
motion to give the full phase-space distortion tensor $\ssvec{D}$
along the trajectory of each simulation particle. This is then
sufficient (through Eq.~(\ref{eq:SymplecticRelations})) to give
all the first order derivatives of the forward and backward
Hamiltonian maps we have been discussing.  In particular, this allows
the local stream density to be calculated at all regular points of 
each trajectory using Eq.~(\ref{eq:StreamDensityRegular}), and
this can be inserted into Eq.~(\ref{eq:AnnihilationRate}) to
obtain the intra-stream annihilation rate at such points.

Caustic crossings can be recognised in an $N$-body integration by the change
in sign of the determinant of $\partial\tvec{x} / \partial \tvec{q}$. Let us
denote by $(s_1,s_2,s_3)$ and $(s_1^\prime,s_2^\prime,s_3^\prime)$ the stretch
factors at the beginning and end of a timestep during which a caustic crossing
is detected. Then $s_1\approx s_1^\prime$ and $s_2\approx s_2^\prime$, while
$s_3$ and $s_3^\prime$ should be much smaller in absolute magnitude and should
have opposite sign. A good approximation to the evolution of the distortion
tensor during the timestep is then obtained by assigning mean values to $s_1$
and $s_2$ and assuming $s_3$ to vary linearly between its endpoint
values. According to the arguments of \S 4.2, the maximum value of $\rho_s$
during the timestep is then well approximated as
\beq 
\rho_{\rm max} =
  \frac{2^{5/4}\Gamma(5/4)~f_0m_p}{\sqrt{\pi}~|\bar{s_1}\bar{s_2}|~
    |\sigma\partial^2{A_3}/\partial{v_3^2}|^{1/2}}.
\label{eq:MaximumStreamDensity}
\eeq
Away from caustic crossing (but still within the timestep) 
Eq.~(\ref{eq:StreamDensityRegular}) can be used to obtain $\rho_s$ which
then varies inversely with $s_3$ and so with $|t - t_c|$, the time to caustic
crossing. This suggests the following approximation to the variation
of stream density within the timestep:
\beqn
  \rho_s(\tvec{x}_p(t)) = \frac{f_0m_p~\Delta
    t}{|\bar{s_1}\bar{s_2}|~|s_3 - s_3^\prime|}~~\lb( (t -
  t_c)^2 + T^2\rb)^{-1/2},
\eeqn
where $\Delta t$ is the length of the timestep and $T$ is chosen so
that $\rho_s(\tvec{x}_p(t_c)) = \rho_{\rm max}$. The shape of this
function in the neighborhood of its maximum is chosen purely for
convenience, but the wings and the maximum value itself are correct.
If we integrate Eq.~(\ref{eq:AnnihilationRate}) over the
timestep using this formula we obtain
\beq
\Delta P = \frac{\langle\sigma_\times v\rangle ~f_0~\Delta
  t}{|\bar{s_1}\bar{s_2}|~|s_3 - s_3^\prime|}~\ln \lb(\frac{2^{9/2}\Gamma(5/4)^2|s_3s_3^\prime|}{\pi\sigma |\partial^2{A_3}/\partial{v_3^2}|}\rb),
\label{eq:TimestepEmission}
\eeq
where we have used the fact that $T\ll \Delta t$ and, for
simplicity, we have assumed that the caustic occurs well away
from either end of the timestep. Since a single simulation particle
represents many dark matter particles, $\Delta P$ must be
multiplied by $m_{\rm sim}/m_p$ (where $m_{\rm sim}$ is the simulation particle
mass) to obtain the total number of annihilation events.

The properties of the dark matter appear in Eq.~(\ref{eq:TimestepEmission})
through the annihilation cross-section, and through the parameter $\sigma$
which appears in the argument of the logarithm.  Thus we recover the result of
\cite{2001PhRvD..64f3515H} that in the cold limit the annihilation luminosity
of a caustic is logarithmically divergent. For given particle physics
parameters, an integration of the $N$-body and geodesic deviation equations
provides all the quantities needed to calculate $\Delta P$ for each caustic
passage, with the important exception of the second derivative
$\partial^2{A_3}/\partial{v_3^2}$. Since this quantity appears only in the
argument of the logarithm, it is sufficient to estimate it to order of
magnitude, provided the estimator chosen has no strong bias when averaged over
many caustic passages.

The condition for caustic passage, ${\rm
  det}(\partial\tvec{A}/\partial\tvec{v}) = 0$ places no constraint on the
values of the second derivatives, so we can estimate the size of a typical
component of $\partial^2\tvec{A}/\partial\tvec{v}^2$ from the available
Galilean-invariant quantities. These are the components of the 6-dimensional
distortion tensor $\ssvec{D}$ and the particle acceleration $\tvec{a}$.  For
example, we can note that $\tvec{a}(\partial^2\tvec{A}/\partial\tvec{v}^2)
(\partial\tvec{x}/\partial\tvec{A})$ is dimensionless and is expected to be of
order unity, so that the size of our desired component of the second
derivative can be estimated as the inverse of the product of the r.m.s. sizes
of the components of $\tvec{a}$ and of the components of
$\partial\tvec{x}/\partial\tvec{A}$. This assumption can be checked in simple
1-dimensional cases, and we will give an example based on the similarity
solution for spherical collisionless collapse in a future paper.  In this
case, at least, the estimate we advocate here works remarkably well.

\section{Discussion}

In this paper we have developed the mathematical background to enable
a relatively precise evaluation of the annihilation radiation from
dark matter caustics in fully general simulations of the nonlinear
growth of structure.  Our scheme allows the annihilation rate to be
integrated along the trajectory of each simulation particle, including
correctly the contributions from all the caustics in which it
participates. Typically each particle experiences several such caustic
passages in each orbit around the dark matter halo in which it resides
\citep[see][]{2008MNRAS.385..236V}. In order to include correctly the
annihilation rate between particles which are members of {\it
  different} streams, it is necessary to estimate a local
coarse-grained density at the position of each particle, and to add in
the contribution due to streams other than its own. This can be done,
for example, using the SPH technique, since the smoothing this
introduces does not bias the luminosity predicted from inter-stream
annihilations. When a particle passes through a caustic occurring in a
stream other than its own, the time-integrated annihilation
probability is still correctly reproduced in the smoothed system.

By implementing these methods in high-resolution simulations of galaxy
formation it should be possible to achieve a complete numerical description of
the expected annihilation radiation, limited only by the ability to resolve
the smallest collapsed clumps of dark matter. This latter limitation can be
severe when attempting to predict the total annihilation radiation from a
representative cosmological volume. For a standard supersymmetric neutralino,
for example, the emission should be dominated by the smallest collapsed
objects, with masses well below that of the Sun
\citep[e.g.][]{2003MNRAS.339..505T}. Recent work has shown, however, that this
problem is much less severe when predicting the observability of annihilation
radiation from the Solar System, which lies just 8 kpc from the centre of the
Milky Way \citep{2008arXiv0809.0894S}. According to these authors, less than 3
percent of the dark matter within 100 kpc of the Galactic Centre should be in
small lumps; almost all the rest should be in extended streams of the kind
discussed in this paper \citep[see also][]{1999MNRAS.307..495H}. They argue that
the highest signal-to-noise for detecting the annihilation signal will be that
of the smooth dark matter distribution in the inner few kiloparsecs of the
Galaxy; small subhalos will be significantly more difficult to
detect. Application of the techniques we have presented here should allow a
rigorous evaluation of an important and previously unresolved issue: whether
the emission structure of this smooth component is significantly modified by
caustic emission. In addition, they will allow an assessment of the expected
morphology and observability of outer caustics around external galaxies, most
notably the Andromeda nebula.

The authors thank St\'ephane Colombi, Craig Hogan, Roya Mohayaee and
Volker Springel for helpful discussions.

\label{lastpage}
\end{document}